\documentclass[9pt]{article}
\usepackage{spconf,amsmath,graphicx}

\usepackage[utf8]{inputenc} 
\usepackage[T1]{fontenc}    
\usepackage{hyperref}       
\usepackage{url}            
\usepackage{booktabs}       
\usepackage{amsfonts}       
\usepackage{nicefrac}       
\usepackage{microtype}      
\usepackage{xcolor}         

\usepackage{graphicx} 
\usepackage{caption}
\usepackage{subcaption}
\usepackage{makecell}
\usepackage{amsmath}
\usepackage{xcolor}         
\usepackage{multirow}

\title{Augmenting Molecular Deep Generative Models\\with Topological Data Analysis Representations}

\name{\parbox{\textwidth}{\centering
    Yair Schiff, Vijil Chenthamarakshan, Samuel C.~Hoffman, Karthikeyan Natesan Ramamurthy, Payel Das
  }
  \thanks{Authors contributed equally.}
}
\address{IBM Research, T.J. Watson Research Center, Yorktown Heights, NY USA}

\begin{document}

\maketitle

\begin{abstract}
Deep generative models have emerged as a powerful tool for learning useful molecular representations and designing novel molecules with desired properties, with applications in drug discovery and material design.
However, most existing deep generative models are restricted due to lack of spatial information.
Here we propose augmentation of deep generative models with topological data analysis (TDA) representations, known as persistence images, for robust encoding of 3D molecular geometry.
We show that the TDA augmentation of a character-based Variational Auto-Encoder (VAE) outperforms state-of-the-art generative neural nets in accurately modeling the structural composition of the QM9 benchmark.
Generated molecules are valid, novel, and diverse, while exhibiting distinct electronic property distribution, namely higher sample population with small HOMO-LUMO gap.
These results demonstrate that TDA features indeed provide crucial geometric signal for learning abstract structures, which is non-trivial for existing generative models operating on string, graph, or 3D point sets to capture.
\end{abstract}

\begin{keywords}
Deep generative modeling,
variational auto-encoder,
molecular design,
topological data analysis
\end{keywords}

\section{Introduction}\label{sec:intro}
Design of new materials is challenging, as it involves exhaustive exploration of the space comprised of small organic molecules on the order of magnitude of 10$^{60}$.
Deep generative models, such as Variational Auto-Encoders (VAEs) and Generative Adversarial Nets (GANs), have emerged as a promising direction for both undirected and goal-directed molecule generation.
These models operate on a variety of inputs, with the popular ones being 1D strings and 2D graphs.
However, these inputs are restricted in terms of capturing 3D structural information, which is known to be crucial for determining chemical properties.
On the other hand, existing generative neural nets that utilize 3D positions of atoms suffer from other limitations, such as lacking a latent representation of molecules, requiring specialized architecture for spatial information modeling, needing an input graph for generation, or sampling nodes (atoms) and edges (bonds) in a less efficient step-wise manner.

In this work, we propose to enhance the spatial information content of a molecular generative model by encoding topological data analysis (TDA) representations.
Using noise-robust topological summaries of molecules, TDA can provide necessary 3D geometric information and can incorporate the influence of other node-level features, such as charge (referred hereafter by the symbol $q$) \cite{carlsson2009theory}.
TDA representations have low computational complexity, since they can be pre-computed.
At the same time, TDA representations are a natural fit for molecular generative models for several other reasons: They incorporate global topological information \cite{carlsson2009topology}, which cannot be captured by molecular graphs or point clouds.
TDA representations are naturally invariant to translations and rotations of molecules and are also equivariant to scaling of distances between atoms of a molecule and permutation of the node order in the molecular graph
\cite{carlsson2009topology} (assuming that topological summaries are derived for functions that do not depend on node order, such as distance between atoms).
Thus, TDA representations are robust to the coordinate system used to represent the molecular parameters.
They are not overly sensitive to noise in the input parameters \cite{cohen2007stability}, which can help in generalization in real-world scenarios where molecules exhibit conformational dynamics.
Features extracted from the distance matrix or graph laplacians of the point cloud representation of atoms do not capture higher level topological features, may not provide scale equivariance and invariance to node order, and have no principled way of incorporating multiple functions, e.g., inter-atom distances and charges.

Our contributions are as follows:
1) We present a principled framework for augmenting molecular deep generative models with TDA representations.
2) Experiments on the challenging QM9 benchmark show that TDA augmentation of a SMILES VAE outperforms existing generative baselines, that are graph-based or 3D point set-based, in modeling structural statistics of the training data.
3) The generated set, when compared to the training set, is valid, novel, and diverse, while also richer in molecules with small HOMO-LUMO gap, a useful property for designing organic photovolatics.

\section{Background}\label{sec:background}
\textbf{Molecular generative auto-encoders}
In recent years, there has been a plethora of work on deep generative models for automatic molecule design.
One line of work \cite{gomez2018automatic, chenthamarakshan2020target, kang2018conditional} considers textual representations of molecules, such as SMILES.
Another strategy has been to utilize molecular graphs that include bond connectivity \cite{simonovsky2018graphvae, jin2018junction}. 
VAE \cite{kingma2013auto} is a popular generative encoder-decoder architecture trained to ensure smoothness on the encoder latent space.
For an input $\mathbf{x} \in \mathbb{R}^d,$ the output of the encoder is a parametric latent distribution $q_{\phi}(\mathbf{z}|\mathbf{x}).$
The output of the decoder is a parameterized posterior distribution on the same space as the input $\mathbf{x}$, $p_{\theta}(\mathbf{x}|\mathbf{z}).$
The VAE loss objective contains a reconstruction term that maximizes the likelihood of the posterior distribution $p_{\theta}$ with a cross-entropy (CE) and a Kullback-Leibler (KL) divergence term, $D_{\textup{KL}}$, between the encoder distribution $q_{\phi}$ and some prior distribution $p(\mathbf{z}),$ which maintains smoothness of the latent $\mathbf{z}$ space.
See the first two terms in Equation \ref{eq:vae_tda}, where $\beta$ is a hyperparameter that controls the weight placed on the KL divergence term.

\noindent\textbf{Persistence images}
We use TDA  to derive representations capturing 3D molecular geometry as well as the charge distributions of the atoms in the molecule.
The representations we use are based on the machinery of persistent homology \cite{edelsbrunner2000topological}.
Given a metric space $X$ and a function $f: X \rightarrow \mathbb{R}$, persistence diagrams characterize the changing topology of the sublevel sets of this function given by $F_{\tau} = {x \in X: f(x) \leq \tau}$ for the parameter $\tau \in \mathbb{R}$.
As the parameter $\tau$ varies, a nested sequence of subspaces is generated, which is referred to as a \textit{filtration} with the property, $F_{\tau_i} \subseteq F_{\tau_j}$ if $\tau_i \leq \tau_j$.
In our case, $X$ is the set of atoms in the molecule.
These atoms are embedded in 3D space.
The function $f$ can be a distance function or a charge function on the atoms.
Persistence diagrams are multisets of points in 2D space, where each point corresponds to the \textit{birth} and \textit{death} parameter values $\tau$ of the homology groups of a given dimension.
In the case of the distance function, the persistence diagrams for dimension $0, 1, 2$ quantify the connected components, holes, and voids in the molecule.
In order to simplify the process of performing statistics on persistence diagrams, several vector-space representations have been proposed \cite{adams2017persistence, bubenik2015statistical}.
Here, we use persistence images \cite{adams2017persistence}, where persistence diagrams are blurred using a Gaussian kernel and samples are drawn from this blurring to produce an image.

\section{Related work}\label{sec:related_work}
\textbf{TDA in molecular/biological systems}
TDA representations have a rich history in analyzing biological systems \cite{camara2017topological}.
For example, both persistent homology and the \textit{Mapper} method \cite{singh2007topological} have been used to extract simple features from Omics data.
Most of those efforts have revolved around predicting properties \cite{krishnapriyan2021machine}, molecular screening \cite{keller2018persistent, Townsend2020RepresentationOM}, or data visualization \cite{topaz2015topological}.
In terms of generative modeling, only recently has there been exploration in using topological priors to regularize the generation of point cloud data \cite{gabrielsson2020topology}.
However, the above works lack solutions for incorporating multiple atomic (intra or inter) parameters toward robust molecular generation.

\noindent\textbf{Encoding 3D molecular geometry in generative models}
Encoding inter-atomic distance matrices has been considered in \cite{hoffmann2019generating} and \cite{nesterov20203Dmolnet}, which do not directly model the 3D geometry. 
SchNet architecture \cite{schutt2018schnet} uses continuous-filter convolutional layers to model inter-atomic spatial interactions and has been used for generating rotational invariant 3D point sets in an autoregressive manner in the G-SchNet framework
\cite{NEURIPS2019_a4d8e2a7}.
NeVAE \cite{JMLR:v21:19-671} utilizes a VAE to learn latent representations of the nodes in a given graph \cite{JMLR:v21:19-671}, which is then used for step-wise decoding into molecular graphs. In contrast to this work, these models lack a continuous latent representation of molecules.

\section{Methodology}\label{sec:method}

In our work, we focus on VAE architectures as the generative models.
We begin with a \textit{vanilla} VAE, which consists of just a single encoder-decoder pair that takes as input and produces SMILES string molecule representations.
We add TDA molecule representations $\mathbf{y} \in \mathbb{R}^n$ in the form of persistence images, which have their own encoder-decoder networks.
The persistence image embedding is concatenated with that from the SMILES string encoder and this concatenated vector is passed to fully connected linear layers to produce the mean $\mu_{\phi}$ and variance $\sigma_{\phi}^2$ that govern the distribution on the latent space $q_{\phi}$.
Once a latent vector is sampled from this space, it is passed to both the SMILES string decoder and the persistence image decoder, which generate $\hat{\mathbf{x}}$ and $\hat{\mathbf{y}}$, respectively.
We then calculate the $\ell_2$ reconstruction loss between the decoded and original persistence images.
Thus our augmented VAE with TDA representation loss becomes
\begin{equation}\label{eq:vae_tda}
\begin{split}
    &\mathcal{L}_{\textup{VAE}+\textup{TDA}}(\theta, \phi; \beta, \lambda) =
    \mathbb{E}_{p(\mathbf{x})}[\mathbb{E}_{q_{\phi}}[\log p_{\theta}(\mathbf{x}|\mathbf{z})]\\
    &- \beta D_{\mathrm{KL}}(q_{\phi}(\mathbf{z}|\mathbf{x}) || p(\mathbf{z}))]
    + \lambda\mathbb{E}_{p(\mathbf{y})}[||\hat{\mathbf{y}} - \mathbf{y}||_2]
\end{split}
\end{equation}
where $\lambda$ controls how much weight is put on the persistence image $\ell_2$ reconstruction loss term.
This augmented loss is easily extended to multiple persistence images by summing $\ell_2$ reconstruction loss across multiple persistence images.
To create the persistence images $\mathbf{y},$ 
we use 3D coordinates and other atomic feature information from the QM9 dataset \cite{ramakrishnan2014quantum, ruddigkeit2012enumeration}.
Persistence diagrams and images are then created with the Ripser \cite{ctralie2018ripser} and Persim packages from Scikit-TDA \cite{scikittda2019}.
We create persistence images for homology dimensions 0, 1, and 2, and flatten and concatenate these together to form the final persistence image vector for each molecule.

For all of our experiments, we use the seminal QM9 benchmark dataset that contains about 134k molecules across the training and test sets \cite{ramakrishnan2014quantum, ruddigkeit2012enumeration}.
This setting of QM9 benchmark is appropriate for testing the expressivity of the proposed generative framework, as the quantum chemical properties are governed by the 3D spatial coordinates of the low energy molecular conformers.
For all VAE models described below, the dimensionality of the latent $\mathbf{z}$ space is 128.
VAE models are trained for 200 epochs with learning rate of $3\times10^{-4}$, Adam optimizer, and a $\beta$ that ranges from $5\times10^{-5}$ to $1\times10^{-2}$, increasing by a fixed step size each epoch.
To test our TDA augmentation, we compare to a baseline VAE (SMILES).
For the SMILES encoder network, we use unidirectional Gated Recurrent Unit (GRU) \cite{cho2014learning} architectures with one layer, hidden dimensions of 256, and dropout of 0.5.
For the decoder, we use the same GRU architecture, with 3 layers, hidden dimensionality of 512, and no dropout.

In augmenting the baseline VAE, we experiment with two combinations of persistence images: 1)  Encoding and decoding persistence images for 3D atomic coordinates (3D) alone and 2) separately encoding and decoding two distinct persistence images, one for atomic 3D coordinates and one for atomic charge (3D + $q$).
Both versions share a common encoder/decoder architecture: 2 fully connected hidden layers with ReLU activation and batch normalization and dropout applied to the first hidden layer.
The final layer of the encoders and decoders is also fully connected.
We use $\lambda =$ 1,000.
Of note, we observed smaller CE reconstruction loss for the TDA-augmented VAEs compared to the baseline VAE.

\section{Results}\label{sec:resuts}
\begin{table*}[t]
    \centering
    \ninept
    \begin{tabular}{l|c||ccc||ccccc}
    & QM9 & SMILES & 3D & 3D + $q$ & GVAE$^*$ & CGVAE$^\dagger$ & MPGVAE$^*$  & MolGAN$^*$ & G-SchNet$^\dagger$ \\
    \midrule
    Validity & 1.000 & 0.819 & 0.840 & 0.852 & 0.810 & 
    \bf{1.000} & 0.91 & \underline{0.98} & 0.771\\
    \midrule    
    \multicolumn{5}{l}{\textit{Atomic composition}}\\
    F & 0.025 & 0.033 & \bf0.019 & \underline{0.018} & 0.235 & -- & 0.127 & -- & --\\
    O &  1.404 & \bf1.406 & 1.295 & 1.303 & 1.017 & 1.528 & \underline{1.457} & 0.861 & 1.786\\
    N & 1.044 & 1.308 & 1.243 & 1.235 & \underline{0.998} & 1.111 & 0.675  & 0.469 & \bf1.071\\
    C & 6.323 & 6.041 & \underline{6.273} & \bf6.282 & 6.750 & 6.898 & 6.740 & 7.454 & 6.064\\
    \midrule\midrule
    Sum & 8.796 & \bf8.789 & \underline{8.829} & 8.837 & 9.000 & -- & 9.000 & -- & --\\
    $\chi^2$ & -- & 0.002 & \bf0.001 & \bf0.001 & 0.014 & -- & 0.009 & -- & --\\
    Sum (No F) & 8.771 & 8.755 & 8.810 & 8.819 & \bf8.765 & 9.537 & 8.872 & \underline{8.784} & 8.921\\
    $\chi^2$ (No F) &  -- & 0.002 & \underline{0.001} & \underline{0.001} & 0.004 & \bf0.000 & 0.005 & 0.025 & 0.003\\
    \midrule
    \multicolumn{5}{l}{\textit{Ring size}}\\
    R3 & 0.470 & 0.479 & \underline{0.462} & \bf0.470 & 0.560 & 0.430 & 0.552 & 0.385 & 0.623\\
    R4 & 0.586 & 0.490 & \underline{0.561} & \bf0.582 & 0.333 & 0.692 & 0.647 & 0.247 & 0.657\\
    R5 & 0.495 & 0.409 & \underline{0.482} & \bf0.483 & 0.218 & 0.902 & 0.526 & 0.325 & 0.430\\
    R6 & 0.158 & 0.169 & \underline{0.155} & \bf0.157 & 0.110 & 0.649 & 0.104 & 0.115 & 0.133\\
    \midrule\midrule
    Sum & 1.709 & 1.600 & \bf1.731 & \underline{1.734} & 1.222 & 2.673 & 1.828 & 1.072 & 1.843\\
    $\chi^2$ & -- & 0.003 & \bf0.000 & \bf0.000 & 0.040 & 0.056 & 0.005 & 0.017 & 0.008\\
    \bottomrule
    \end{tabular}
    \caption{Chemical validity, atomic composition, and ring size distribution of 
    generated molecules. Sum total and $\chi^2$ distances between QM9 ground truth and generated histograms are also provided for compositional analyses. Number next to ``R'' refers to the ring size.
    Best and second best results are indicated by bold and underline, respectively. 
    Baseline values taken from \cite{Flam_Shepherd_2021} = *,
     \cite{NEURIPS2019_a4d8e2a7} = $\dagger$.
    }
    \label{tab:counts}
\end{table*}

\noindent\textbf{Accuracy of generated molecules} 
We compare the chemical validity, along with the structural statistics, such as atom composition and ring structure count, of the generated molecules from our VAE variants to the QM9 molecules and to those generated from existing generative baselines\footnote{Atom and ring counts for baseline models were extracted from figures in \cite{NEURIPS2019_a4d8e2a7} and \cite{Flam_Shepherd_2021} using \url{https://github.com/ankitrohatgi/WebPlotDigitizer} \cite{Rohatgi2019}.}.  
The following state-of-the-art generative models are considered as baselines: three graph-based VAEs -- GraphVAE \cite{simonovsky2018graphvae} (GVAE, unconditional NoGM version), a graph VAE that incorporates valency constraints (CGVAE) \cite{liu2018constrained}, and a message passing graph VAE (MPGVAE) \cite{Flam_Shepherd_2021} -- and a graph-based GAN (MolGAN) \cite{de2018molgan}.
We also compare against the G-SchNet generative model that leverages 3D geometry \cite{NEURIPS2019_a4d8e2a7}.
Chemical validity (fraction of \textit{total} generated molecules that are valid, as determined by the RDKit library \cite{landrum2013rdkit}) is estimated on 30k molecules randomly sampled from the   prior. 
 Table \ref{tab:counts} shows that the TDA augmentation encourages valid molecule generation (compared to the SMILES-only model), and results are    on par with the existing baselines, particularly the ones that do not impose additional reward \cite{de2018molgan} or constraint \cite{liu2018constrained} to promote validity. Structural compositions are then estimated, after removing duplicates, non-novel molecules, and non-valid SMILES.
We also provide sum total and chi-squared distances for compositions; $\chi^2(\mathcal{X}, \mathcal{X'}) = \frac{1}{2}\sum_{i=1}^{n}\frac{(x_i - x'_i)^2}{(x_i + x'_i)}$ (where $x_i$ is the normalized value of category $i$ in $\mathcal{X}$, i.e. $x_i =  y_i / \sum_{j=1}^{n}y_j$), between histograms of QM9 and generated set in Table \ref{tab:counts}.
Smaller $\chi^2$ values, combined with better approximation to QM9 mean, show that the TDA-augmented VAEs more accurately capture both atom  and ring counts than all baselines, including G-SchNet that utilizes 3D coordinates in learning \cite{NEURIPS2019_a4d8e2a7}.
Interestingly, some of the baseline models, e.g. CGVAE and our SMILES-only VAE, perform relatively poorly in preserving the structural statistics, particularly ring structures.
These results demonstrate that the TDA features indeed provide crucial geometric signal for learning abstract structures, such as loops and voids, which is not trivial for character or graph VAEs to capture.
Further, TDA representations appear to be a robust means for encoding global geometry, when compared to existing baseline like G-SchNet.

\begin{table*}
    \ninept
    \centering
    \begin{tabular}{l|cc||ccccccc}
        & 3D & 3D + $q$ & CVAE$^*$ & GVAE$^*$ & CGVAE$^{\$}$ &  MPGVAE$^\dagger$ & MolGAN$^\dagger$ & G-SchNet$^{\$}$ & NeVAE\\
        \midrule
        Uniqueness@10k & \underline{0.984} & 0.981 & 0.675 & 0.241 & \bf0.986 & 0.68 & 0.10 & 0.919 & $-$\\
        Novelty & 0.611 & 0.594 & 0.900 & 0.610 & \underline{0.943} & 0.54 & \underline{0.94} & 0.875 & \bf1.000\\
        \bottomrule
    \end{tabular}
    \caption{Uniqueness (for 10k) and novelty of  generated valid molecules. 
    Bold and underline text in each row indicates highest and second highest values, respectively.
    Baseline values taken from \cite{simonovsky2018graphvae} = *, \cite{Flam_Shepherd_2021} = $\dagger$, \cite{NEURIPS2019_a4d8e2a7} = $\$$.
    }
    \label{tab:moses_metrics}
\end{table*}

\noindent\textbf{Quality of generated molecules} 
 We further evaluate the quality of the TDA-VAE generated molecules by estimating their 
uniqueness (fraction of the \textit{valid} strings that are unique) and novelty (fraction of the \textit{valid} strings not present in the training set) and compare those with existing baselines. Results are presented in Table \ref{tab:moses_metrics}, which includes
 a character-based VAE (CVAE) trained on SMILES \cite{gomez2018automatic} and NeVAE \cite{JMLR:v21:19-671} (with no masking) as baselines, in addition to the  ones from Table \ref{tab:counts}. 
We follow the convention of CVAE, GraphVAE, and MPGVAE and report uniqueness as a fraction of 10k generated \textit{valid} molecules.
NeVAE reports uniqueness as a fraction of their total 1M generated molecules (of which 68.2\% are valid), and so we refrain from reporting NeVAE performance on uniqueness for a fair comparison.
Together, Tables \ref{tab:counts} and \ref{tab:moses_metrics} imply that the TDA-augmented VAEs can generate high fraction of unique molecules that are chemically valid and consistent with the training distribution in terms of structural metrics, while showing moderate novelty.    
In contrast, existing baselines are limited in terms of one (or more) of the following: uniqueness, validity, or closeness to QM9 structural characteristics. 

We further investigate  
the capability of the TDA-augmented model in generating molecules with a desired electronic property, such as low  HOMO-LUMO gap (< 5eV) that is crucial for designing organic semiconductors and is sparsely seen in the training molecules.
For this purpose, we relax 5,000 generated novel and valid molecules at the same level of theory as QM9 (B3LYP / 6-31G(2df,p)).
The generated set from the 3D + $q$ model contains 13.75\% molecules with desired HOMO-LUMO gap, when compared to 7.56\% in the training set and 8.67\% in G-Schnet generated molecules.

\noindent\textbf{Latent space characterization}
It is natural to check to what extent the latent space of TDA-VAEs contains 3D geometric signal \cite{schiff2020characterizing}.
To do so, we estimate the correlation between the Euclidean distance in the $\mathbf{z}$ space and the persistence image distance for each pair of QM9 test molecules for the TDA-augmented VAE variants and the baseline VAE.
Significantly higher correlation is observed for the TDA-augmented VAEs (0.376 for the 3D variant and 0.406 for the 3D + $q$ variant), when compared to the baseline VAE (0.214), implying that the learned latent embeddings indeed contain important 3D geometric features. Further, the correlation between the $\mathbf{z}$ distances and the Tanimoto similarities on MACCS keys was found to be higher with TDA augmentation (0.323 for 3D and 0.350 for 3D + $q$) compared to the baseline (0.306).
This result indicates that the  presence of TDA features in the latent representation enhances expressivity in terms of capturing structural information at different levels, including presence of substructures and persistent homology features.

We further check the consistency between persistence images decoded from randomly sampled latent vectors and those extracted post-hoc from decoded SMILES strings of the same randomly sampled vectors.
Using the 3D + $q$ variant on a sample size of 10k, we find errors that are 0.003\% of the 3D coordinate and 0.002\% of the charge average persistence image norms.
This analysis indicates that the latent space of the TDA-augmented VAE encodes enough meaningful and `sensible' information about persistence images and that generating `non-meaningful' persistence images is rare.

Finally, we measure consistency between decoded persistence images and latent embeddings by randomly drawing two points from the latent space of the 3D + $q$ TDA-augmented VAE and linearly interpolating 1,000 equally spaced steps in between these two endpoints.
For the interpolations, we then decode their persistence images.
We compare interpolation step size $t\in[0,1]$ to $\ell_2$ distance between the decoded persistence image of the interpolated point and the decoded persistence image of the endpoint  $t=0$.
By averaging over 100 repeated runs, we find the Pearson correlation between step size and these (averaged) distances is 0.999.
This result implies that the latent space of the TDA-augmented VAE is highly smooth with respect to persistence images.

\section{Conclusion}\label{sec:conclusion}
We have presented a framework for robust encoding of crucial geometric information in a molecular deep generative model by utilizing TDA representations.
Results show, for the first time, that augmenting a VAE trained on SMILES with persistence images can enhance the model's accuracy, in terms of improving validity and better modeling of structural statistics.
The TDA-augmented VAE is able to generate molecules that are novel, unique, and exhibit desired electronic properties sparsely seen in the training data.
Future work will explore the mathematical underpinnings of how TDA representations help encode 3D information into generative latent spaces, the use of TDA augmentation in other architectures and beyond small organic molecules, and direct decoding into 3D structures.
We will also investigate regularization of generative models using topological priors, in the vein of  \cite{gabrielsson2020topology}.

{\fontsize{9.0pt}{10.0pt}\selectfont \bibliographystyle{IEEEtran}
\bibliography{refs}
}

\newpage
\appendix
\onecolumn
\section{Appendix}\label{app}
\subsection{Assets}\label{app:assets}
The QM9 dataset is publicly available for download here: \url{http://quantum-machine.org/datasets}\\

\noindent The packages used in this work and their respective licenses are listed below:
\begin{enumerate}
    \item MOSES; \url{https://github.com/molecularsets/moses}; MIT 
    \item NWChem; Educational Community License (ECL) 2.0
    \item PyMol; Custom (\url{https://github.com/schrodinger/pymol-open-source/blob/master/LICENSE})
    \item PyTorch; BSD
    \item PyTorch Lightning; Apache 2.0
    \item OpenBabel; GPL 2.0
    \item RDKit; BSD
    \item Sciki-TDA (Ripser and Persim); MIT
\end{enumerate}

\subsection{Persistent homology on molecule features}\label{app:persim}
\begin{figure}[ht!]
    \centering
    \includegraphics[width=.25\linewidth]{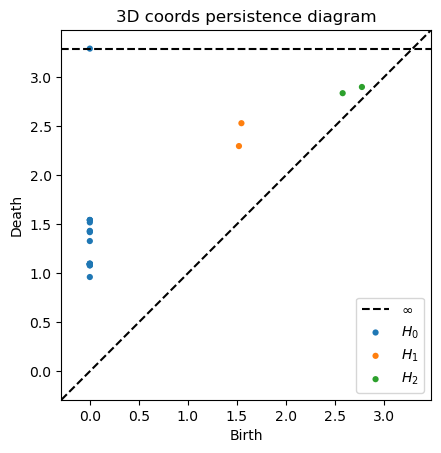}
    \includegraphics[width=.25\linewidth]{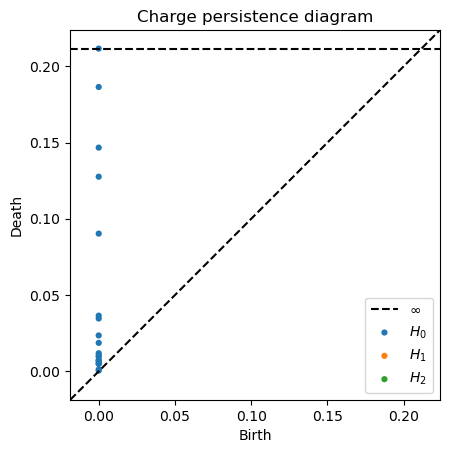}
    \caption{Sample persistence diagram for 3D atomic coordinates (left) and atomic charge (right) for molecule \texttt{OC1CCC2OC1C=C2}.}
    \label{fig:persistence_diagram}
\end{figure}

\begin{figure}
    \centering
    \includegraphics[width=0.7\textwidth]{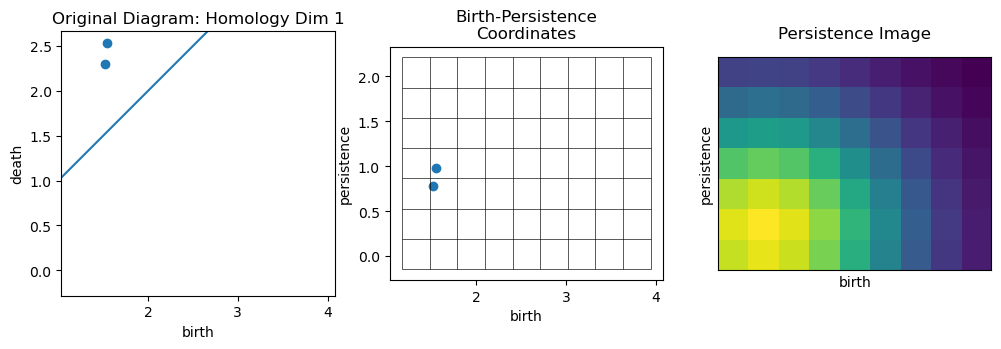}
    \caption{Sample for how to create a persistence image from the 3D coordinates persistence diagram for homology dimension 1 for the molecule with SMILES string \texttt{OC1CCC2OC1C=C2}.}
    \label{fig:persistence_image}
\end{figure}

In Figure \ref{fig:persistence_diagram}, we show a sample persistence diagrams for 3D atomic coordinates and for atomic charge for the molecule with SMILES string \texttt{OC1CCC2OC1C=C2}.
In Figure \ref{fig:persistence_image}, we show an example of how to go from the 3D coordinate persistence diagram for homology dimension 1 to a persistence image for this same molecule.

\subsection{Additional VAE experimental setup}\label{app:vae_experiment}
For QM9 VAE training, we split the data such that the training set contains about 121k and the test set contains about 13k molecules.
For the VAE training paradigms below, we use a compute environment with 1 CPU and 1 V100 GPU and submit training as 6 hour jobs to a cluster.
Sampling is done in a compute environment with 1 CPU submitted as 6 hour jobs to a cluster.
Below, we describe the persistence image, encoder, embedding, and decoder dimensions for the TDA-augmented VAE models.
All persistence image dimensions refer to flattened and concatenated (across homology dimensions) vectors.\\

\noindent\textbf{3D coords}
For the VAE augmented with a single persistence image for 3D coordinates, the persistence image dimension is 102, the encoder hidden dimensions are 64 and 32, the embedding dimension is 16, and the decoder hidden dimensions are 32 and 64.\\

\noindent\textbf{3D coords + $q$}
For the VAE augmented with separate persistence images for 3D coordinates and charge, the persistence image dimension for 3D coords is 102, the 3D coord encoder hidden dimensions are 64 and 32, the embedding dimension is 16, and the decoder hidden dimensions are 32 and 64.
The persistence image dimension for charge is 28, the charge encoder hidden dimensions are 16 and 8, the embedding dimension is 4, and the decoder hidden dimensions are 8 and 16.

\subsection{Generated molecule visualization}
\begin{figure}[ht!]
    \captionsetup[subfigure]{labelformat=empty,justification=centering}
    \centering
    \begin{subfigure}[b]{0.18\textwidth}
    \centering
    \includegraphics[width=\textwidth]{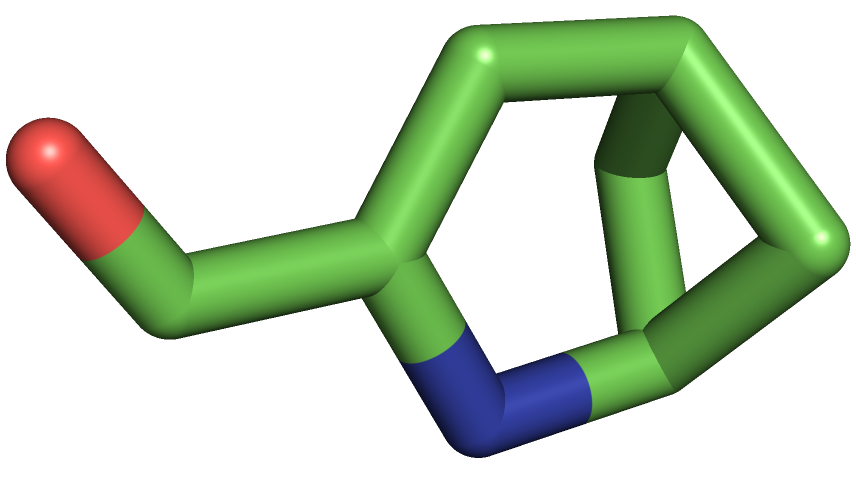}
    \includegraphics[width=\textwidth]{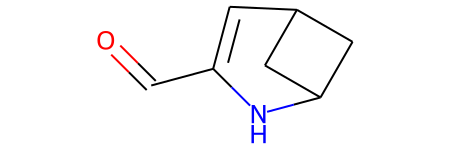}
    \caption{\texttt{O=CC1=CC2CC(C2)N1}\\Gap $=$ 4.226 eV}
    \end{subfigure}
    \hfill
    \begin{subfigure}[b]{0.18\textwidth}
    \centering
    \includegraphics[width=\textwidth]{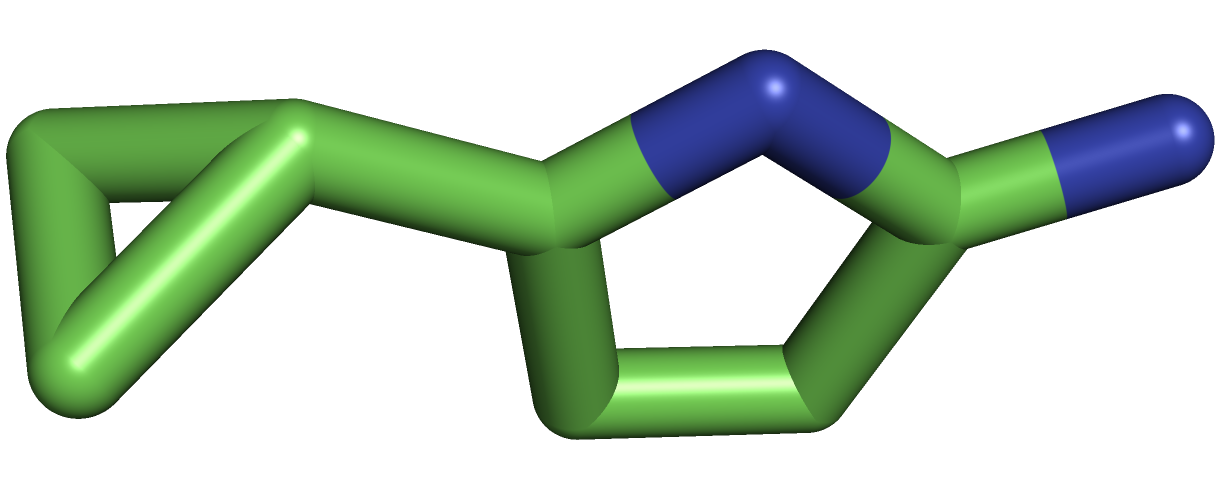}
    \includegraphics[width=\textwidth]{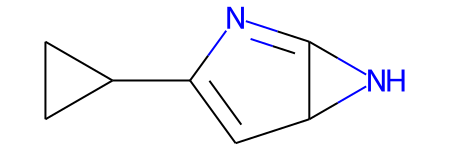}
    \caption{\texttt{C1=C(C2CC2)N=C2NC12}\\Gap $=$ 4.635 eV}
    \end{subfigure}
    \hfill
    \begin{subfigure}[b]{0.18\textwidth}
    \centering
    \includegraphics[width=\textwidth]{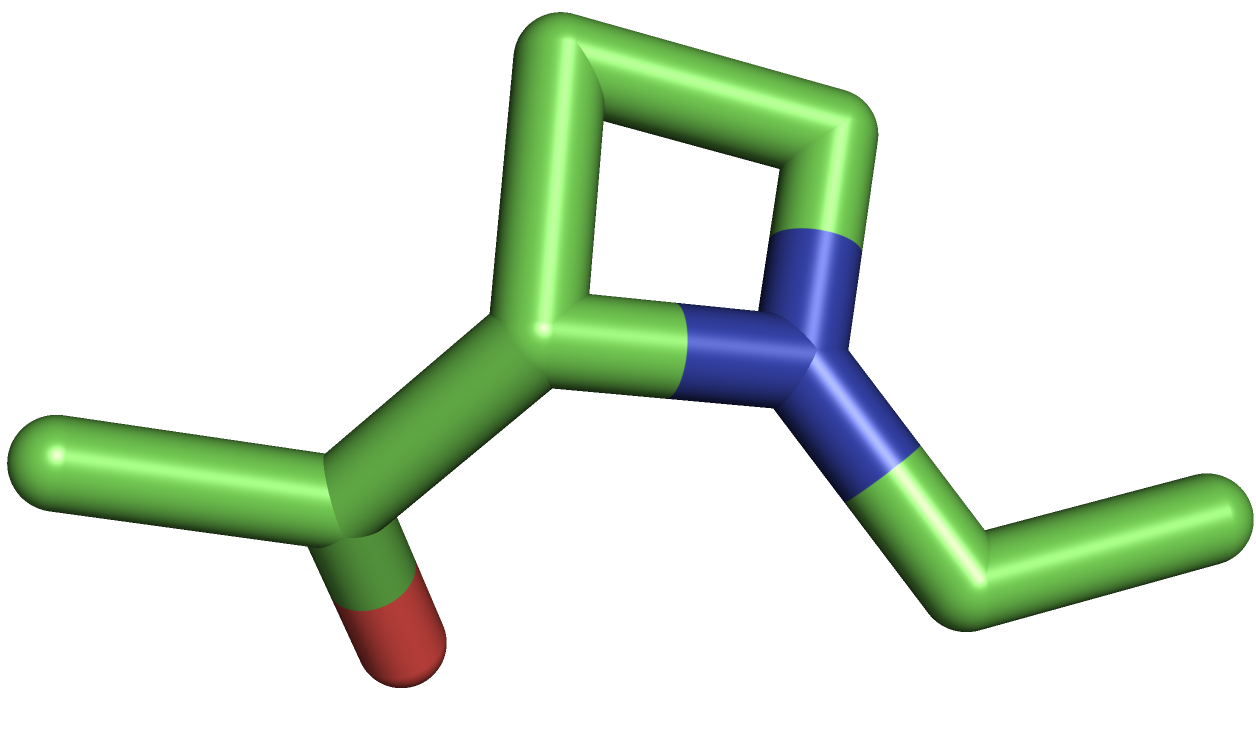}
    \includegraphics[width=\textwidth]{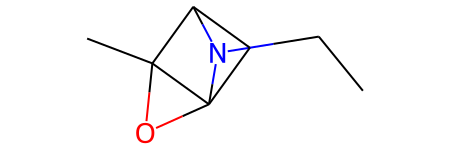}
    \caption{\texttt{CCN1C2CC13OC23C}\\Gap $=$ 4.555 eV}
    \end{subfigure}
    \hfill
    \begin{subfigure}[b]{0.18\textwidth}
    \centering
    \includegraphics[width=\textwidth]{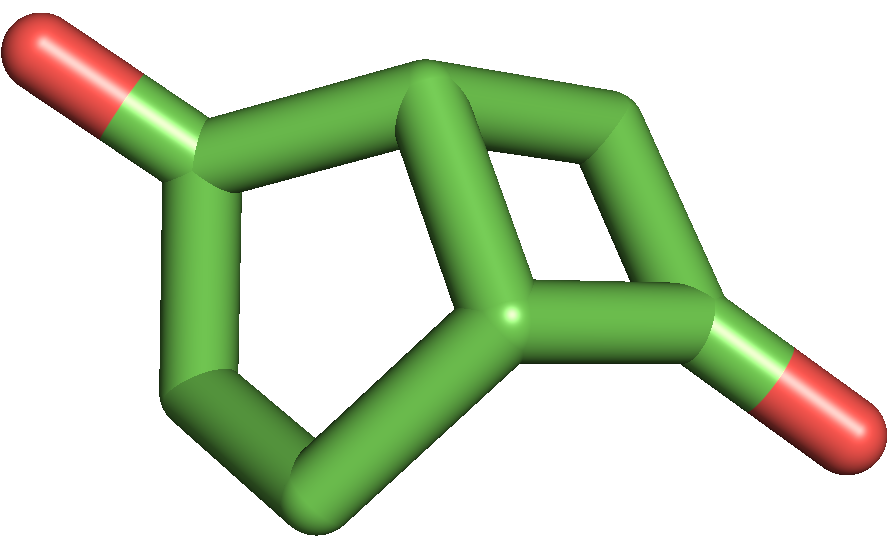}
    \includegraphics[width=\textwidth]{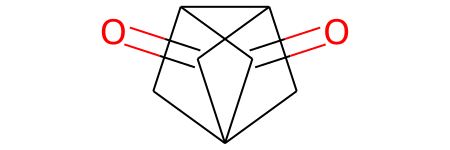}
    \caption{\texttt{O=C1C2CC13CC2C3=O}\\Gap $=$ 2.840 eV}
    \end{subfigure}
    \hfill
    \begin{subfigure}[b]{0.18\textwidth}
    \centering
    \includegraphics[width=\textwidth]{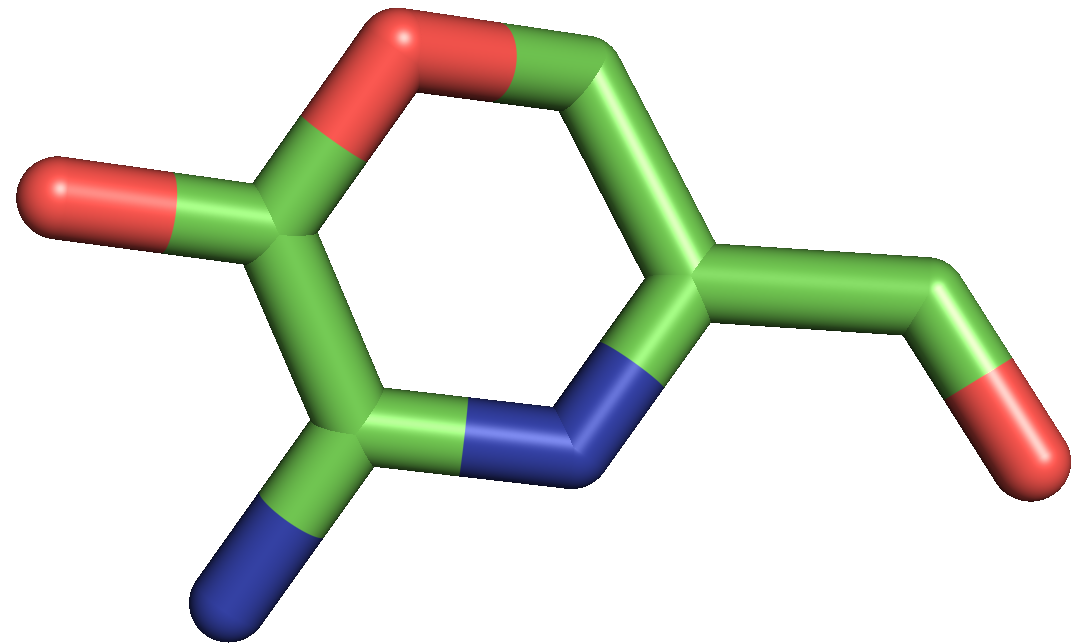}
    \includegraphics[width=\textwidth]{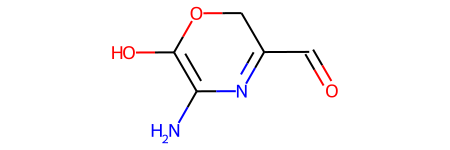}
    \caption{\texttt{NC1=C(O)OCC(C=O)=N1}\\Gap $=$ 2.990 eV}
    \end{subfigure}
    
    \vspace{20pt}
    
    \begin{subfigure}[b]{0.18\textwidth}
    \centering
    \includegraphics[width=\textwidth]{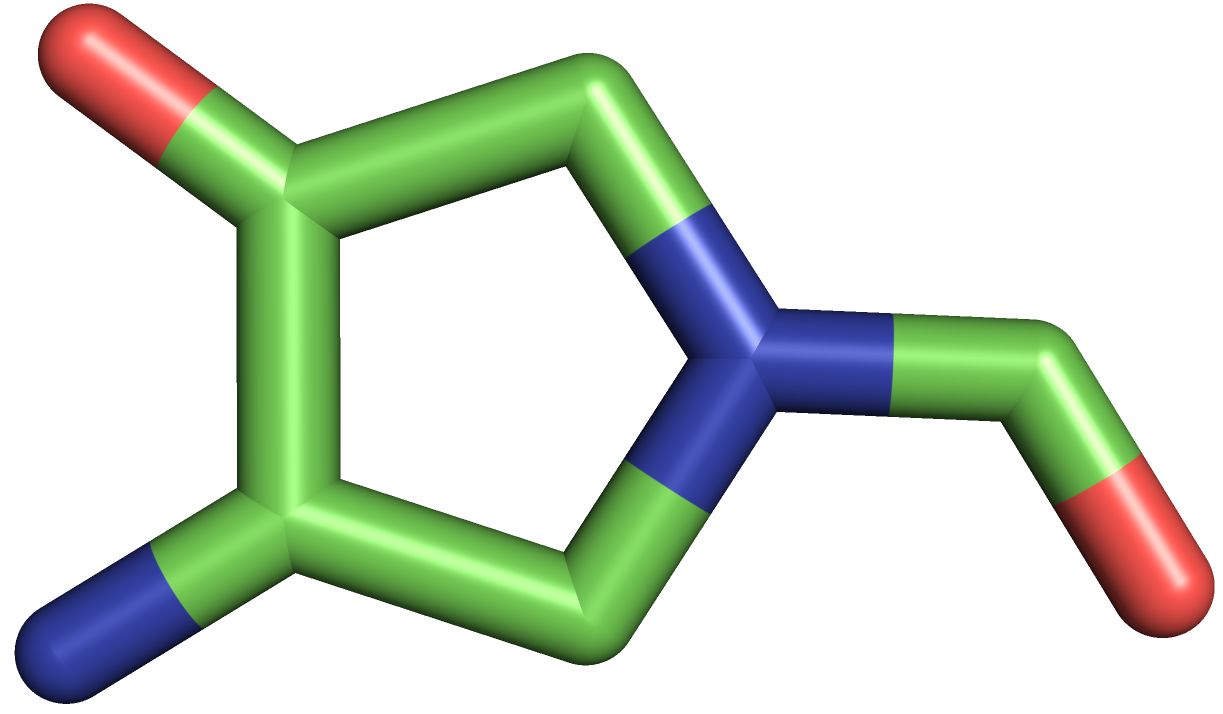}
    \includegraphics[width=\textwidth]{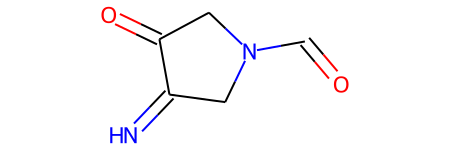}
    \caption{\texttt{N=C1CN(C=O)CC1=O}\\Gap $=$ 4.585 eV}
    \end{subfigure}
    \begin{subfigure}[b]{0.18\textwidth}
    \centering
    \includegraphics[width=\textwidth]{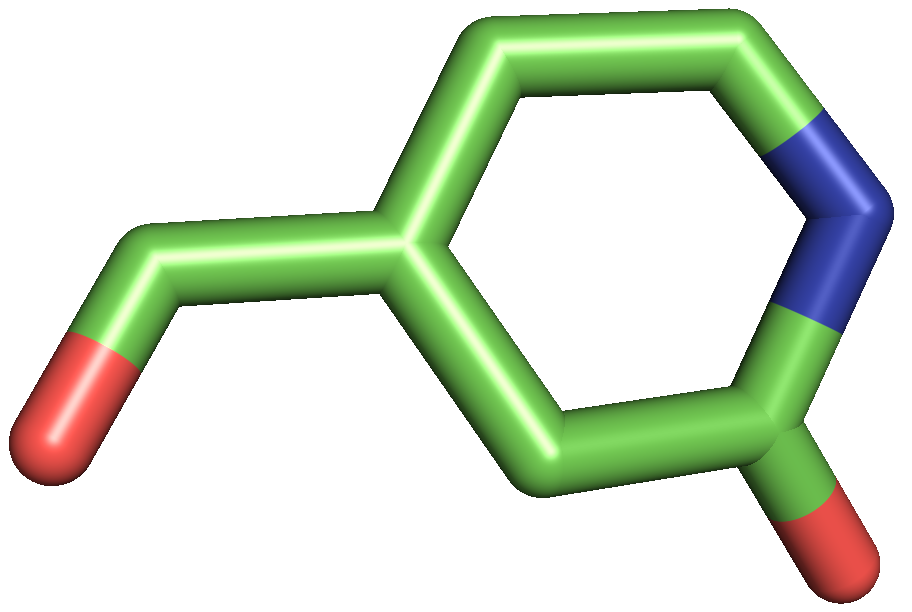}
    \includegraphics[width=\textwidth]{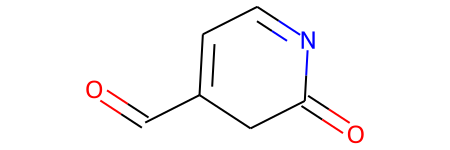}
    \caption{\texttt{O=CC1=CC=NC(=O)C1}\\Gap $=$ 3.791 eV}
    \end{subfigure}
    \begin{subfigure}[b]{0.18\textwidth}
    \centering
    \includegraphics[width=\textwidth]{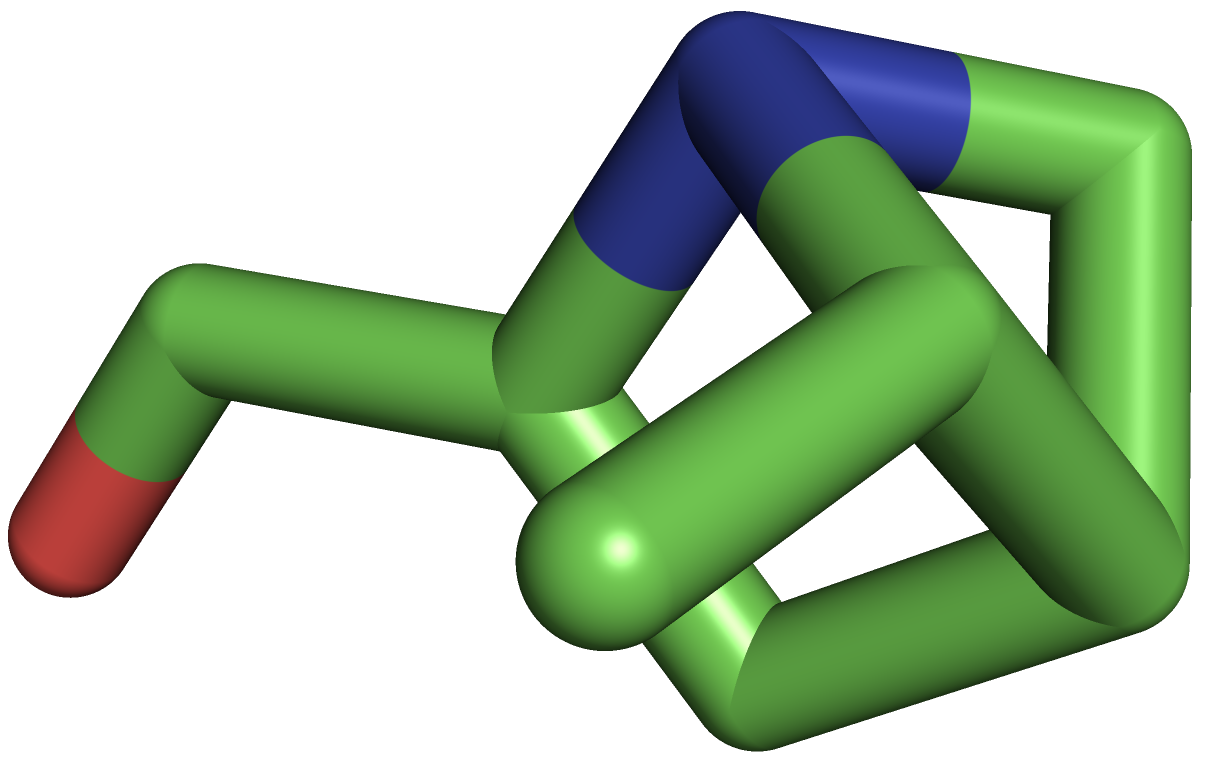}
    \includegraphics[width=\textwidth]{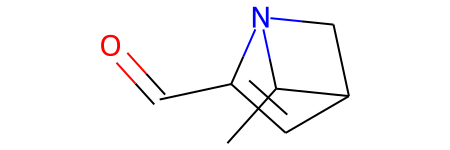}
    \caption{\texttt{CC1C2C=C(C=O)N1C2}\\Gap $=$ 4.885 eV}
    \end{subfigure}
    \begin{subfigure}[b]{0.18\textwidth}
    \centering
    \includegraphics[width=\textwidth]{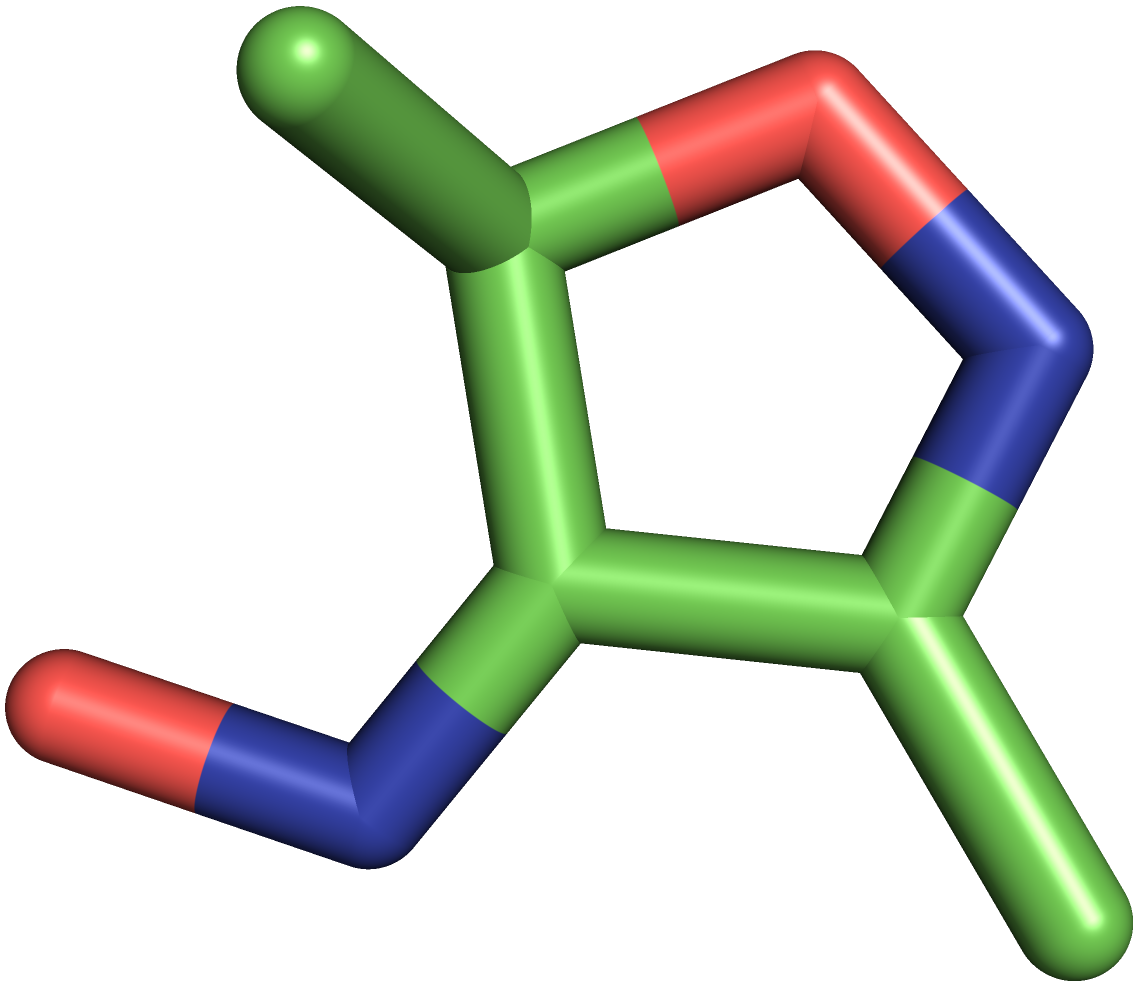}
    \includegraphics[width=\textwidth]{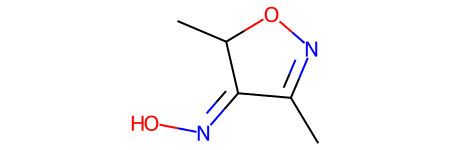}
    \caption{\texttt{CC1=NOC(C)C1=NO}\\Gap $=$ 4.974 eV}
    \end{subfigure}
    \begin{subfigure}[b]{0.18\textwidth}
    \centering
    \includegraphics[width=\textwidth]{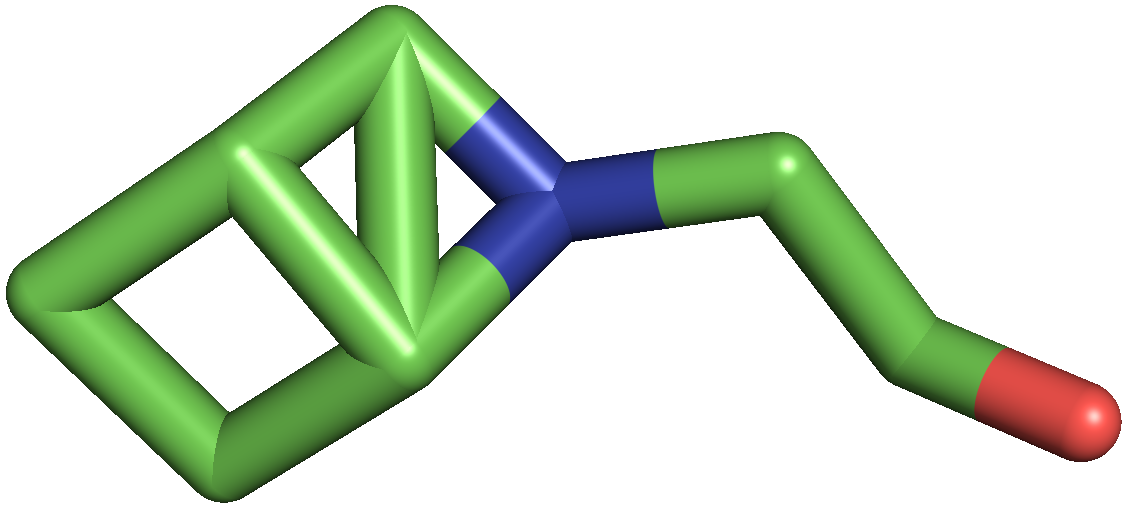}
    \includegraphics[width=\textwidth]{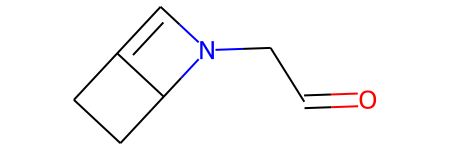}
    \caption{\texttt{O=CCN1C=C2CCC21}\\Gap $=$ 4.869 eV}
    \end{subfigure}

    \caption{SMILES, 2D graphs, and relaxed 3D geometry from NWChem for 10 randomly generated molecules from the 3D coords + $q$ TDA-VAE with  HOMO-LUMO gaps < 5eV.}
    \label{fig:low_gap_mols}
\end{figure}

In Figure \ref{fig:low_gap_mols}, we show SMILES, 2D graph, and relaxed 3D geometry for 10 randomly chosen generated molecules from the 3D coords + $q$ variant of TDA-VAE with estimated HOMO-LUMO gap < 5 eV.
QM calculations were each run on a compute cluster with 8 CPUs in parallel with MPI for up to 6 hours with the following parameters:
\textit{task:} DFT;
\textit{basis:} 6-31g(2df,p) (obtained from the Basis Set Exchange -- \url{https://www.basissetexchange.org}); \textit{xc:} B3LYP;
\textit{grid:} fine;
\textit{iterations:} 120;
\textit{optimizer:} driver, 120 iterations.

\end{document}